\documentstyle[aps,twocolumn,epsf]{revtex}      
\begin{document}
\title{Distribution of level curvatures for the
Anderson model at the localization-delocalization transition} 
\author{C.~M.~Canali$^1$, Chaitali~Basu$^1$, W.~Stephan$^2$ 
and V.~E.~Kravtsov$^{1,3}$}
\address{ $^1$ International Centre for Theoretical Physics, 34100 Trieste, 
Italy \\
$^2$Dipartimento di Fisica, Universit\`a "La Sapienza", Roma \\
$^3$Landau Institute for Theoretical Physics,
Moscow, Russia
\\[3pt]
(Received 20 December 1995)
%
\\ \medskip}\author{\small\parbox{14cm}{\small                             
We compute the distribution function of single-level curvatures, $P(k)$,
for a tight binding model with site disorder, on a cubic lattice. 
In metals $P(k)$
is very close to the predictions of the random-matrix theory (RMT).
In insulators $P(k)$ has a logarithmically-normal form.
At the Anderson localization-delocalization transition
$P(k)$ fits very well the proposed novel distribution
$P(k)\propto (1+k^{\mu})^{3/\mu}$ with $\mu \approx 1.58$,
which approaches the RMT result for large $k$
and is non-analytical at small $k$.
We ascribe
such a non-analiticity to the spatial multifractality of
the critical wave functions.
\\[3pt]PACS numbers: 71.30.+h, 
72.15.Rn,                                    
05.60.+w}}\address{}\maketitle  

An important consequence of the one-parameter scaling theory\cite{1PS}
of the
Anderson transition is the existence of three
universality classes for the energy level statistics of disordered
systems. In the metallic regime they are described by the
strongly correlated Wigner-Dyson (WD) distribution, while in the
insulating regime they follow an uncorrelated Poissonic law.
This difference has its fundamental origin
in the underlying nature of the corresponding eigenstates,
being extended and strongly overlapping in the first case
but localized in the second.

The properties of the third universal statistics, describing the
spectral correlations at the critical point, have been only
recently the subject of intense investigation, both analytical
and numerical.
Using the results of numerical simulations,
Shklovskii {\it et al.}\cite{ShSh} suggested that the spacing distribution
function $P(s)$ has the WD form $P(s)\sim s$ for small $s$ and
the Poissonic tail $P(s)\sim e^{-s}$ for large $s$. Further analytical
investigations\cite{KLAA,AKL} showed that the two-level correlation
function $R(s)$ in the critical region has a novel power-law asymptotic decay
$R(s)= - c/s^{2-\gamma}$ with a nontrivial exponent $\gamma=1-1/\nu d$.
Here $\nu$ is the critical exponent of 
the correlation/localization length $\xi$,
which depends on the dimensionality $d$.
Thus the two-level correlation function
in the critical region resembles qualitatively the WD function which
applies to the metallic phase.
On the other hand, the level number variance $\Sigma_{2}(N)=\langle
(\delta N)^{2} \rangle$ in an energy strip of width $N\Delta$,
($\Delta$ is the mean level spacing),
still contains a dominant 
Poissonic term,\cite{altshlinear88,Sears,BraunMontambaux}
linear in $N$, which is typical
for insulators. 

It can be shown\cite{KLAA} that the Poissonic term in
$\Sigma_{2}(N)$ is only
possible if a normalization sum rule on $R(s)$ is violated 
in the thermodynamic limit (TL) $L\rightarrow \infty$,
($L$ is the system size)
which signals a qualitative change in the statistics of wave functions
$\Psi({\bf r})$.
An analogous situation occurs in certain
random matrix ensembles with broken unitary symmetry, that turn out to
describe the critical statistics $\Sigma_{2}(N)$ and $P(s)$ 
very well.\cite{cmcK}

Since the Poisson statistics describe localized states, it was
put forward \cite{KravtsovNC} that
the cause of the linear term in $\Sigma_{2}(N)$ is the existence
near the Anderson transition of pre-localized states, for which
sharp peaks in $|\Psi({\bf r})|^{2}$
contribute significantly to the normalization integral.
In fact the existence of the pre-localized states - even in good
metals - had been already conjectured in Ref.~\cite{Habilitationsschrift}
for the interpretation of a slow current
relaxation in disordered conductors\cite{AKL91}.
Recently this conjecture has been confirmed
in a more direct way using the supersymmetric 
sigma-model.\cite{Muzykantskii,FalkoEfetov}.
Furthermore, the 
multifractal structure of wave functions $\Psi({\bf r})$, expected at the
Anderson transition,\cite{Wegner}
can be represented\cite{FalkoEfetov}
as a superposition of peaks in $|\Psi({\bf r})|^{2}$
corresponding to pre-localized states with a certain {\it distribution}
of amplitudes and exponents for their power-law spatial decay.

A statistical property that may shed some light on the problem
of pre-localized states
is the distribution of curvatures of the single energy-levels.
The level curvature measures the sensitivity of the energy spectrum
to a change of the boundary conditions. For each single energy
level $\epsilon_n(\varphi)$, the curvature $K_n$ is defined as
\begin{equation}
K_n = {1\over \Delta} 
{\partial^2 \epsilon_n(\varphi)\over \partial\varphi ^2}
\biggr\vert_{\varphi =0}
\end{equation}
The parameter $\varphi$ enters
in the definition of generalized boundary conditions in the $z$ direction
satisfied by the single-level wave functions, 
$\Psi(x,y,z + L) = e^{i2\pi\varphi}\Psi(x,y,z)$. For a quasi-onedimensional
sample, the parameter $\varphi$ is equivalent to the phase generated by
an Aharonov-Bohm flux $\phi = \varphi \phi_0$ that pierces a closed ring.
The way in which an eigenvalue responds to a small twist of the boundary
condition is obviously related to the nature of the corresponding wave
function. Typically an eigenstate extended through the sample will feel
any change in the boundary conditions and will have a large curvature, 
whereas a localized
state, far from the edges, will be insensitive to the twist and its 
curvature will be close to zero. However we will see that this
intuitive argument must be taken with caution.

In this paper we investigate numerically the curvature distribution of the
$3d$ Anderson model. Similar calculations have been performed earlier
by Zyczkowski {\it et.~al.},\cite{molinari}, 
who studied the distribution for all the three regimes of the disorder, 
and by Braun and
Montambaux\cite{braunandmont},
who just looked at the metallic (diffusive) regime.
The purpose of our work is
to carry out an accurate scale analysis of the full curvature
distribution right at the {\it metal-insulator transition}, 
which has not been done
so far. We will first compute very accurately the 
curvature distribution for the metallic and insulating regime, and
provide a physical interpretation of the numerical results
in these two cases, based on the nature of their wave functions. 
We will then show that
the critical distribution can be fitted extremely well by a
functional expression similar to the RMT result that applies
to metals, except for the presence of a new nontrivial exponent,
which makes the distribution nonanalytical at $K=0$.
We will finally argue that such a behavior support the hypothesis of
pre-localized states in the critical region.

We consider a tight binding model on a cubic lattice of
volume $V= (L)^3$, where $L$ is
is the number of sites in any direction. 
The one-particle Hamiltonian is
\begin{equation}
H= \sum_i \epsilon_i c_i^\dagger c_i + 
t\sum_{\langle i j\rangle}(e^{i\theta_{ij}} c^\dagger _i c_j + 
e^{-i\theta_{ij}}c^\dagger _j c_i)
\end{equation}
The site energies $\epsilon_i$ are randomly distributed 
with uniform probability
between $-W/2$ and $+W/2$. The parameter $W$ controls the amount of 
disorder in the system. The phase shifts $\theta_{ij}$ in the hopping term 
are chosen
in such a way that when a particle hops from $z$ to $z+ L$ it picks up
a total phase shift equal to $e^{i 2\pi \varphi}$ coming from generalized
boundary conditions in the $z$ direction.
The Anderson transition is known to take
place in such a system for $W_c/t\approx 16.5$ when $\varphi =0$.

For each disorder configuration we can compute the zero-flux curvature of 
each single level. Since in the presence of the disorder the level velocity
$v_n = {\partial \epsilon_n\over \partial \varphi}\bigr\vert_{\varphi =0}$
is zero, the curvature
can be calculated numerically in terms of the following finite difference
\begin{equation}
K_n(\varphi =0)= 2{[\epsilon_n(\varphi) - \epsilon_n(0)]\over {\varphi^2}}
\end{equation}
In order
to get accurate approximations of the differentials with finite differences,
particular care must be taken in the choice of the 
$\varphi$.\cite{molinari,braunandmont}
We fitted the energy shift $[\epsilon_n(\pm \varphi) - \epsilon_n(0)]$
to a parabola $K_n\varphi^2$ and checked that the value found
for $K_n$ is the same when we double $ \varphi$.
The optimal value of $\varphi$ ranges form $10^{-1}$
to $10^{-4}$ 
depending on disorder and system size, increasing with disorder.
Our ensemble averages typically include several thousands disorder
realizations, each one retaining half of the spectrum
in an energy-window centered at $\epsilon=0$..

We start with the metallic case where we know that classical RMT is 
applicable. Delande and Zakrewski\cite{ZDp}
conjectured that
the curvature distribution for complex systems described by RMT
follows the expression 
\begin{equation}
\label{ZD}
{\cal P}_{\beta}(k) = C_{\beta}{1\over (1+ k^2)^{1 + \beta/2}}
\end{equation}
with $\beta = 1,2,4$ for the orthogonal, unitary and symplectic
ensemble respectively. The constant $C_{\beta}$ is fixed by normalization.
The dimensionless curvature $k$ is given by
$k= K/\langle K\rangle$, where $\langle K\rangle$ is the first
moment of the distribution.
It has been shown quite recently\cite{vonoppen,FS}
that Eq.~(\ref{ZD}) is in fact exact for ensembles
of large Gaussian matrices for all three symmetries.

\begin{figure}
\epsfxsize=4.5truecm \epsffile{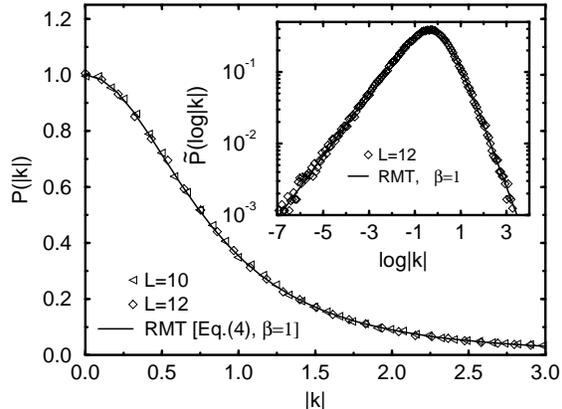}
\caption{Distribution of curvatures in the metallic regime $W/t=5$,
for different $L$, plotted with
Eq.~(\protect\ref{ZD}) for $\beta=1$ as a comparison. 
In the inset  ${\tilde P}(\log|k|)=|k|P(|k|)$ vs $\log|k|$.}
\label{fig1}
\end{figure}

In Fig.~\ref{fig1} we plot our results for the distribution
of the rescaled curvatures $P(k)$ for several system sizes and
$W=5t< W_c$. 
It is seen that the RMT expression, Eq.~(\ref{ZD}) with $\beta =1$,
is remarkably
accurate in the metallic regime for the whole range of $k$.
It is universal, in the sense that,
after rescaling $K$, $P(k)$ is scale invariant and 
disorder independent.
Our results
agree with Refs.~\cite{molinari,braunandmont}
and settle the question of the $1/k^3$ dependence of $P(k)$ at 
large $k$, implying the divergence of the second moment.
We emphasize that, for the $3d$ Anderson model in the metallic regime,
$P(k)$ agrees very
well with ${\cal P}_{\beta}(k)$ of Eq.~(\ref{ZD}) {\it also near} $k=0$, 
in contrast to the
case of other quantum chaotic
systems\cite{ZDp} that display nonuniversal features at small 
$k$, due to
{\it scarring} of their wave functions. This point will be very useful
in the interpretation of the critical $P(k)$ near $k=0$.

An important property of ${\cal P}_{\beta}(k)$
is the maximum at $k=0$.
Intuitively, on the basis of the above arguments, one would expect
that in the metallic phase, where the eigenfunctions are extended,
most of the curvatures should be large. In fact, as we show below,
the maximum in $P(k)$ is related to the ``ergodic'' nature or
lack of specific internal structure of the typical metallic eigenstate.
Using second-order perturbation theory,
the second derivative of the eigenvalues with respect to the parameter
$\varphi$ due to a small perturbation is written as
\begin{equation}
\label{PT}
{\partial^2 \epsilon_n\over \partial \varphi^2}\biggr\vert_{\varphi =0}
\propto \sum_{m \neq n} {\bigl\vert\langle m|J_z|n\rangle\bigr\vert^2
\over  \epsilon_m(0) - \epsilon_n(0)}\; ,
\end{equation}
where $|m\rangle$ and $\epsilon_m(0)$ are the eigenvectors and eigenvalues
of $H$ at zero flux, respectively. $J_z$ is the current operator
in the $z$ direction.
If we assume an ordered spectrum 
$\epsilon_{m}-\epsilon_{n}=\Delta(m-n)$,
then for each term of this sum relative to the eigenvalue 
$\epsilon_{m_+}$, separated from $\epsilon_n$ by an energy shift 
$+\delta \epsilon_n = \epsilon_{m_+} - \epsilon_n$, 
there will be another term corresponding to $\epsilon_{m_-}$
such that $\epsilon_{m_-} - \epsilon_n =-\delta \epsilon_n$.
Moreover the matrix elements of these two terms $|\langle {m_+}|J_z|n\rangle|$
and $|\langle {m_-}|J_z|n\rangle|$ must be equal if the corresponding
unperturbed eigenstates are extended and structureless. Therefore we
conclude that there is a cancellation in the sum of Eq.~(\ref{PT}),
causing the curvature to be zero. Of course the actual distribution is
not simply $\delta(k)$ because the spectrum is not fully ordered.
Large values of the curvature $|k|\sim 1/s $ are due to anomalously small
spacings 
$s=|\epsilon_{m}(0)-\epsilon_{n}(0)|/\Delta \ll 1$ for $m=n+1$ or $m=n-1$,
which occur with probability $P(s)\sim s^{\beta}$. Thus for large
$|k|$ one obtains a power-law decay ${\cal P}_{\beta}(k)\sim
\int ds\, s^{\beta}\,\delta(|k|-1/s)\sim |k|^{-(2+\beta)}$ instead of zero,
but the maximum of ${\cal P}_{\beta}(k)$ remains at $k=0$.

\begin{figure}
\epsfxsize=4.5truecm \epsffile{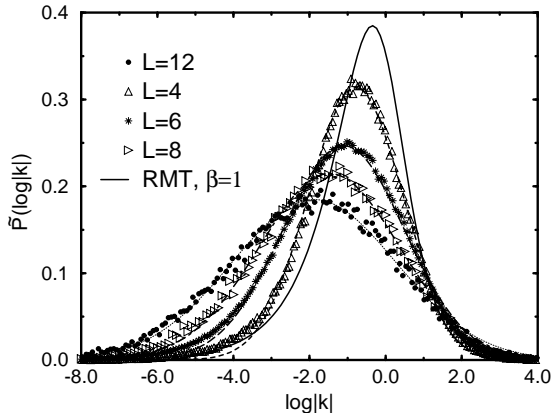}
\caption{Distribution of curvatures ${\tilde P}(\log|k|)$ vs $\log|k|$
for the insulating regime $W/t =30$ for different
$L$'s. The dotted, dashed, etc. lines are
gaussian fittings. The solid line is the RMT result for $\beta=1$ 
in these coordinates.}
\label{fig2}
\end{figure}

This simple argument allows us to
interpret also the results for the other extreme case, namely the
regime of strong disorder where localization is present.
Indeed, from Eq.~(\ref{PT})
we see that now the cancellation in the sum
does not occur: two states $| {m_+}\rangle$ and 
$| {m_-}\rangle$, separated in energy from $| n\rangle$ by
$\pm \delta \epsilon$ will have matrix elments
that differ greatly from each other,
since they are, in general, localized at different points.
Therefore for a large but finite
system size $L$, the typical level curvature is small but
{\it non-zero}. Thus $P(k\rightarrow 0) =0$ for an insulator.
We can guess the form of the distribution function in this case
if we assume that the curvature $k$ is related to the amplitude
of the typical wave function at the edges, $\log|k| \sim \log |\Psi|$,
with $|\Psi| \sim \exp(-L/L_o)$, where $L_{o}$ is the radius of the
localized wave function.
If the distribution of $1/L_{o}$ is 
a Gaussian 
$P(1/L_o) \propto e^{-({1\over L_o} - {1\over \xi})^2}$
centered around the inverse localization length $\xi^{-1}$,
we come 
immediately to the conclusion that the curvature distribution for an insulator
must be log-normal, $P(|k|) = C\exp[-A(\log|k| + B)^2]$, with
$A\sim 1/L^{2}$ and $B\sim L/\xi$. The numerical
results of the simulations for the insulating regime
are shown in Fig.~\ref{fig2}, where we plot
${\tilde P}(\log|k|)=|k|P(|k|)$ vs. $\log|k|$ for $W =30t > W_c$ 
and different system sizes.
In these coordinates the distribution can be fitted rather well by gaussians,
when the system size is large ($L=12$),
implying that $P(k)$ is log-normal, in agreement with what is found
in Ref.~\cite{molinari}. The vanishing of P(k) at $k=0$
occurs very close to the origin and therefore is not easily resolved
in a $P(k)$ vs $k$ plot.  However for smaller $L\le 8$, 
${\tilde P}(\log |k|)$ clearly deviates from a gaussian distribution
at small $k$
and approaches the RMT function even for such a large disorder.

\begin{figure}
\epsfxsize=4.5truecm \epsffile{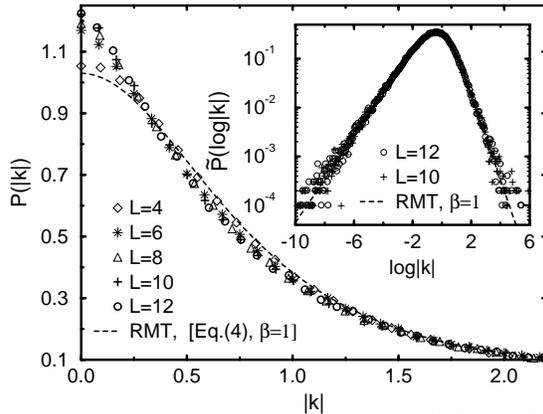}
\caption{Distribution of curvatures $P(|k|)$ vs $|k|$
at the critical point ($W_c/t=16.5$) for different
$L$'s. In the inset ${\tilde P}(log|k|)$ vs $\log|k|$ for $L=10,12$.}
\label{fig3}
\end{figure}

\begin{figure}
\epsfxsize=4.5truecm \epsffile{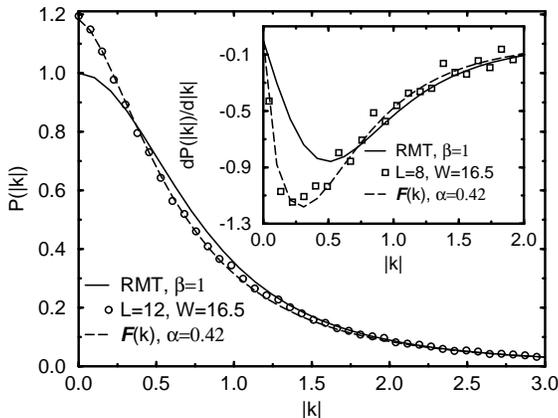}
\caption{$P(|k|)$ vs. $|k|$ at the critical point
for $L=12$. The dashed line is the fitting function of
Eq.~(\protect\ref{fitf}) with $\alpha=0.42$. The solid line is the
RMT function, Eq.~(\protect\ref{ZD}), $\beta=1$. In the inset
$dP(|k|)/d|k|$ calculated numerically from the $P(|k|)$ data for $L=8$,
together with the derivatives of ${\cal F}(k)$
and ${\cal P}_{\beta}(k)$.}
\label{fig4}
\end{figure}

We finally come to the central part of our work, the curvature 
distribution in the critical region. It was pointed out
before that $P(k)$ can be used to locate efficiently
the Anderson transition point,\cite{molinari}
since according to the one-parameter scaling
$P(k)$ must be scale invariant at this point. However, there has been so far
no detailed study of the full correlation function at the Anderson transition
and its dependence on the sample size $L$. The results for different system
sizes at $W_{c}/t=16.5$ are shown in Fig.3 together with the RMT result
${\cal P}_{\beta=1}(k)$ for comparison.
One can see that the curvature
distribution is close to Eq.~\ref{ZD} valid for metals. 
In particular, for large
$k$ the two curves are indistinguishable within our numerical precision.
Thus even at the critical point the $P(k)$ follows the
$|k|^{-3}$ rule, which corresponds to a spacing distribution
$P(s) \sim s$.
On the other hand the curvature distribution $P(k)$ at the critical
point deviates slightly but significantly from ${\cal P}_{\beta}(k)$ 
at small $k$. 
Here the maximum at $k=0$ moves above the metallic limit and apparently
is size dependent.
This seems to indicate that, in contrast to the
prediction of one-parameter scaling, even at the transition point
there is a {\it finite} length scale $R$ such as for $L<<R$ the
full distribution is indistinguishable from the RMT result, and for $L>>R$
it saturates to a complete scale-invariant curve that differs from Eq.~\ref{ZD}
at small $k<1$.
As shown in Fig~\ref{fig4}, where we plot the critical $P(k)$ for $L=12$,
this curve fits very well the distribution function
\begin{equation}
\label{fitf}
{\cal F}(k) = {A\over [1 + k^{\mu}]^{3\over \mu}} \; ,
\end{equation}
where $A= \mu\Gamma(3/\mu)[\Gamma(1/\mu)\Gamma(2/\mu)]^{-1}>1$
is determined by normalization.
Once normalized, the distribution in Eq.~\ref{fitf} obeys the condition
$\langle k\rangle =1$ for
{\it all} $\mu= 2 - \alpha$. Thus the full distribution ${\cal F}(k)$
is determined uniquely by its value at $k=0$, and with the known
$P(k=0)$ the fit of the whole curve is parameter free. This procedure
yields $\alpha = 0.42 \pm 0.01$. 
For further evidence we have plotted, in the inset of Fig.~\ref{fig4}, 
the derivative
of $P(k)$ with respect to $k$,
together with the derivative of ${\cal F}(k)$.
Equation~(\ref{fitf}) implies that 
$dP/d|k| \sim |k|^{1-\alpha}$ for small $|k|$. 

Such a non-analiticity in $P(k)$ is the signature of the
complexity of the wave functions at the critical point.
Another (related) feature of the critical distribution $P(k)$
is an enhanced probability to find a level with a small 
value of curvature. It can result from the multifractal spatial structure
of the wave functions in the critical region, provided that there is no
drastic difference
in matrix elements $\langle {m_+}|J_z|n\rangle$
and $\langle {m_-}|J_z|n\rangle$ so that $P(K=0)$ remains nonzero.
Indeed, such a structure
can be considered as a limiting case of ``scarring'' with a 
certain distribution
of sharp peaks in $|\Psi({\bf r})|^{2}$, which exhibits itself in the
spectrum of the multifractal dimensions $d_p<d$ seen 
in the inverse participation ratios (IPR) $I_{p}$:
\begin{equation}
I_{p}\propto \int |\Psi({\bf r})|^{2p} d^{d}{\bf r}\sim L^{-(p-1)d_p}.
\end{equation}
Notice that in a metal $I_{p} \sim 1/L^{(p-1)d}$ and in the
insulator $I_{p} \sim constant$ in the TL. Thus one can say that the
behavior at the critical point resembles the one in metals, since in both
cases the IPR vanishes when $L\to \infty$. However the exponents
are different. This situation is analogous to what we find for
the curvature distribution.

In conclusion, on the basis of numerical diagonalization of the
tight-binding Anderson model we have suggested a novel level 
curvature distribution function $P(k)$ at the Anderson transition,
which exhibits a non-analytical
behavior for small $k$ and approaches the RMT result for large $k$. We relate
this non-analyticity to the spatial multifractality of critical wave functions.

One of us (C.~M.~C.) would like to thank R.~Hlubina for interesting discussions.


\end{document}